\documentclass [preprint,
 aps,showpacs]{revtex4}
\usepackage[final]{graphics}
\usepackage{amssymb}
\usepackage{amsfonts}
\usepackage{epsfig}
\usepackage{graphicx}
\usepackage{subfigure}

\begin{document}
\title{Contact process on a Voronoi triangulation
}
\author{Marcelo M. de Oliveira $^1$\footnote{corresponding author:
mancebo@fisica.ufmg.br},
 S. G. Alves$^2$, S. C. Ferreira$^1$ and Ronald Dickman$^3$}
\affiliation{$^{1}$Departamento de F\'{\i}sica, Universidade Federal
de Vi\c{c}osa, 36571-000, Vi\c{c}osa, MG, Brazil\\$^{2}$Campus Alto
Paraopeba, Universidade Federal de S\~ao Jo\~ao Del Rei, 36420-000,
Ouro Branco - Minas Gerais, Brazil. \\$^{3}$Departamento de
F\'{\i}sica, ICEx, Universidade Federal de Minas Gerais, 30123-970,
Belo Horizonte - Minas Gerais, Brazil.}

\date{\today}

\begin{abstract}
We study the continuous absorbing-state phase transition in the
contact process on the Voronoi-Delaunay lattice. The Voronoi
construction is a natural way to introduce quenched coordination
disorder in lattice models. We simulate the disordered system using
the quasistationary simulation method and determine its critical
exponents and moment ratios. Our results suggest that the
critical behavior of the disordered system is unchanged with respect to
that on a regular lattice, i.e., that of directed percolation.
\end{abstract}


\maketitle

\section{Introduction}

Nonequilibrium phase transitions between an active (fluctuating)
state and an inactive, absorbing state arise frequently in
interacting particle models \cite{marro}, chemical catalysis
\cite{ziff86}, interface growth \cite{tang} , epidemics
\cite{bartlett} and related fields. In spatially extended systems,
exemplified by the contact process \cite{harris-CP}, such
transitions are currently of great interest, which has been
heightened by recent experimental confirmations of absorbing-state
phase transitions in a liquid crystal system \cite{take07}, and in a
sheared colloidal suspension \cite{pine}. Much of this work is
focused on issues of universality, aimed at characterizing the
critical behavior of these models in terms of universality classes
\cite{marro,hinrichsen,odor04,lubeck}. It has been conjectured
\cite{jans81,gras82} that models with a positive one-component order
parameter, short-range interactions, and absence of additional
symmetries or quenched disorder belong generically to the
universality class of directed percolation (DP), which is considered
the most robust universality class of transitions to an absorbing
state.

The contact process (CP) is one of the simplest and most studied
models belonging to the DP universality class. Of particular
interest is how spatially quenched disorder affects its critical
behavior \cite{cafiero}. Quenched disorder, in the form of
impurities and defects, plays an important role in real systems, and
may be responsible for the rarity of experimental realizations of DP
\cite{hinri00b}. Quenched disorder in the contact process on a
regular lattice has been studied in the forms of random deletion of
sites or bonds \cite{noest,adr-dic,vojta06}, and of random spatial
variation of the control parameter \cite{vojta05,durrett,salinas08}.
All these studies report a change in the critical behavior of the
model. These findings are consistent with Harris' criterion
\cite{harris74}, which states that quenched disorder is a relevant
perturbation if
\begin{equation}
d\nu_\perp<2,
\end{equation}
where $d$ is the dimensionality and $\nu_\perp$ is the correlation
length exponent of the pure model (In DP this inequality is
satisfied in all dimensions $d<4$, since $\nu_\perp =$ 1.096854(4),
0.734(4) and 0.581(5), for $d=1$, $2$ and $3$, respectively
\cite{jensen99,voigt97,jensen92}.) Some controversy remains whether
the exponents change continuously with degree of disorder
\cite{adr-dic,hooy}, or whether they change abruptly to the values
in the strong disorder limit corresponding to the universality class
of the random transverse Ising model, as suggested by Vojta in a
recent work \cite{vojta05}.

Harris' criterion determines the relevance of disorder in the form
of independent random dilution (of sites and/or bonds) in a regular
lattice. A somewhat different situation arises when the underlying
graph is not periodic, as is the case in a deterministic aperiodic
structure, or in a graph with a random neighbor structure such as
the Voronoi triangulation. To determine the relevance of disorder in
these cases, the following heuristic extension of Harris' criterion
was proposed by Luck \cite{luck93}: Consider a spherical patch
$\Omega$ with radius $R$ on a given realization of a graph. The
patch encloses a number $B(R)$ of vertices, which scales as
$B(R)\sim R^d$. The average coordination number in the patch is
given by
\begin{equation}
J(R)=\frac{1}{B(R)}\sum_{i\in \Omega}q_i,
\end{equation}
Let the fluctuation of the coordination number around its expected
value, $J_o=\overline{q}$, decays as
\begin{equation}
\sigma_R(J)=\frac{\langle \mid J(R)-J_o \mid
\rangle}{J_o}\sim\langle B(R)\rangle^{-(1-\omega)}\sim
R^{-d(1-\omega)},
\end{equation}
when $R\to \infty$. Here, $\omega$ is defined as the {\it wandering
exponent} of the triangulation.  Nearby the critical point
$\Delta\equiv(\lambda-\lambda_c)/\lambda_c=0$, the fluctuations
$\sigma_\xi(J)$ of the average coordination number in a correlation
volume scale as
\begin{equation}
\sigma_\xi(J)\sim\xi_\perp^{-d/2}\sim\Delta^{\nu_\perp d/2},
\end{equation}
since $\xi_\perp\sim \Delta^{-\nu_\perp}$. Considering a large
correlation volume, $R\sim\xi_\perp$, the resulting shift of the
critical point, induced by the fluctuations $\sigma_\xi$ in a
correlation volume is proportional to
$\Delta^{d\nu_\perp(1-\omega)}\sqrt{\mbox{var}(q_i)}$, where
$\mbox{var}(q_i)=\langle q_i^2 \rangle-\langle q_i \rangle^2$. Then,
in order that the regular critical behavior remain unchanged,
these fluctuations should die out when $\Delta\to 0$, which is true
if $\omega$ does not exceed a threshold value given by
\begin{equation}
\omega_c=1-\frac{1}{d\nu_\perp}.
\end{equation}
Thus, in principle, the Harris-Luck criterion permits one to predict
the effects of quenched disorder in models defined on structures
such as quasi-crystals or even random lattices. (Note that for
independent dilution, $\omega=1/2$, and Luck's expression reduces to
the Harris criterion.)

In this work we investigate whether disorder in the form of a
quenched Poissonian {\em coordination disorder} alters the critical
behavior of the contact process, by studying the critical behavior
of the process on a Voronoi-Delaunay (VD) type random lattice
\cite{okabe,hilh08}. The VD lattice represents a natural way of
introducing quenched coordination disorder in a lattice model, and
also plays an important role in the description of idealized
statistical geometries such as planar cellular structures, soap
throats, etc. \cite{okabe}. In this lattice, the sites are spatially
distributed following a Poisson distribution, and the coordination
number $q$ varies randomly,  with $3 \leq q < \infty$ and
$\overline{q} = 6$ in the infinite-size limit. Our results suggest
that coordination disorder does not change the critical behavior of
the contact process.

The balance of this paper is organized as follows. In the next
section we review the definition of the contact process and detail
construction of the VD lattices as well the simulation methods used.
In Sec. III we present our results and discussion; Sec. IV is
devoted to our conclusions.

\section{Model and Method}

Consider a bounded domain $\Omega$ in a $d$-dimensional space in
which $N$ nodes are randomly placed with uniform distribution. The
Voronoi diagram of this set is a sub-division of the domain into
regions $V_i$ (with $i = 1, 2, \ldots, N$), such that any point in
$V_i$ is closer to node $i$ than to any other node in the set.
Figure 1 (a) shows a patch of a Voronoi diagram. The points whose
cells share an edge are considered neighbors. The dual lattice,
obtained by linking neighboring sites is the Voronoi-Delaunay
network, exemplified in Fig.1 (b). One of the characteristics of the
dual lattice is that its local coordination number varies randomly,
with the distribution shown in Fig.2. In this work, we take periodic
boundary conditions,{\it i.e}, the domain $\Omega$ has a toroidal
topology. In order to construct the lattices we follow the method of
Ref. \cite{frie-ren}. For simplicity, we express the length $L$ of
the domain $\Omega$ in terms of the size of a regular lattice
$L=\sqrt{N}$.

\begin{figure}[h]
\begin{center}
\includegraphics[width=8cm,clip=true]{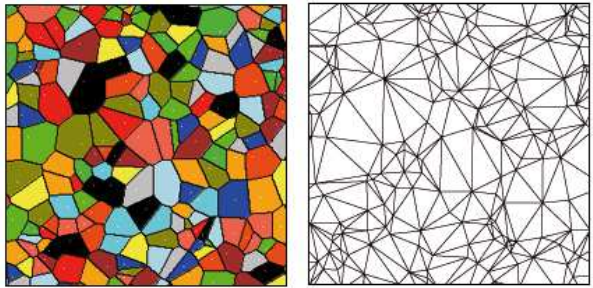}
\end{center}
\caption{(a) A patch of a Voronoi Diagram. (b) The corresponding
dual lattice to the diagram shown in (a).(Color
  online).}
\end{figure}

\begin{figure}[!htb]
\begin{center}
\includegraphics[width=8cm,clip=true]{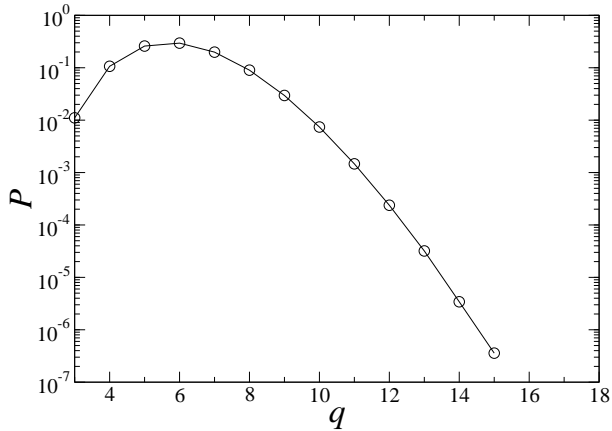}
\end{center}
\caption{Degree distribution $P(q)$ of the Voronoi-Delaunay lattice,
for system size $L=2560$.}
\end{figure}

The CP, originally introduced as a ``toy model" for epidemic
spreading \cite{harris-CP}, is a stochastic interacting particle
system defined on a lattice, with each site either occupied
($\sigma_i (t)= 1$), or vacant ($\sigma_i (t)= 0$). Transitions from
$\sigma_i = 1$ to $\sigma_i = 0$ occur at a rate of unity,
independent of the neighboring sites. The reverse transition is only
possible if at least one of its neighbors is occupied: the
transition from $\sigma_i = 0$ to $\sigma_i = 1$ occurs at rate
$\lambda r$, where $r$ is the fraction of nearest neighbors of site
$i$ that are occupied; thus the state $\sigma_i = 0$ for all $i$ is
absorbing. ($\lambda $ is a control parameter governing the rate of
spread of activity.)

In the simulation we employ the usual simulation scheme
\cite{marro}, in which annihilation events are chosen with
probability $1/(1+\lambda)$ and creation with probability
$\lambda/(1+\lambda)$. In order to improve efficiency, the sites are
chosen from a list of currently occupied sites. In the case of
annihilation, the chosen site is vacated, while, for creation
events, one of its $q$ nearest-neighbor sites is selected at random
and, if it is currently vacant, it becomes occupied.  The time
increment associated with each such event is $\Delta t = 1/N_{occ}$,
where $N_{occ}$ is the number of occupied sites just prior to the
attempted transition.

In the studies reported here we sample the {\it quasistationary}
(QS) distribution of the process, (that is, conditioned on
survival), which has proven a very useful tool in the study of
processes with an absorbing state \cite{marro,prue07,dickman08}. For
this purpose, we employ a simulation method that yields
quasistationary (QS) properties directly, the QS simulation method
\cite{qssim}. The method is based in maintaining, and gradually
updating, a set of configurations visited during the evolution; when
a transition to the absorbing state is imminent the system is
instead placed in one of the saved configurations. Otherwise the
evolution is exactly that of a conventional simulation.

\section{Results and Discussion}

We performed extensive simulations of the CP on Voronoi-Delaunay
random lattices of $L=20,40,...,640$, using the QS simulation
method. Each realization of the process is initialized with all
sites occupied, and runs for at least $10^8$ time steps. Averages
are taken in the QS regime, after discarding an initial transient
which depends on the system size. This procedure is repeated for
each realization of disorder (For each size studied,  we performed
averages over 200-300 different lattices).

\begin{figure}[htb]
\begin{center}
\includegraphics[width=8cm,clip=true]{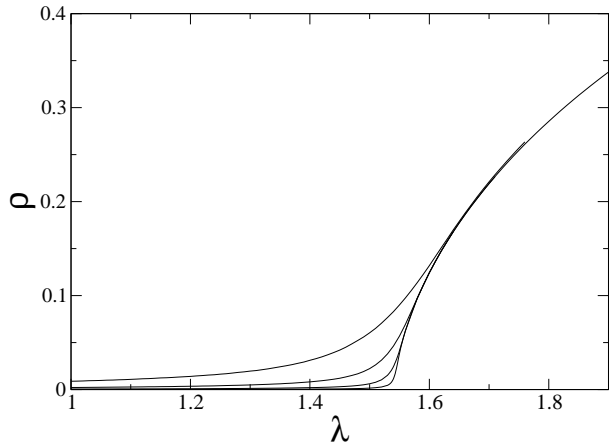}
\end{center}
\caption{Quasistationary density of active sites $\rho$ as a
function of the control parameter $\lambda$. System sizes: $L=20$,
40, 80 and 160, from top to bottom.}
\end{figure}

\begin{figure}[!htb]
\begin{center}
\includegraphics[width=8cm,clip=true]{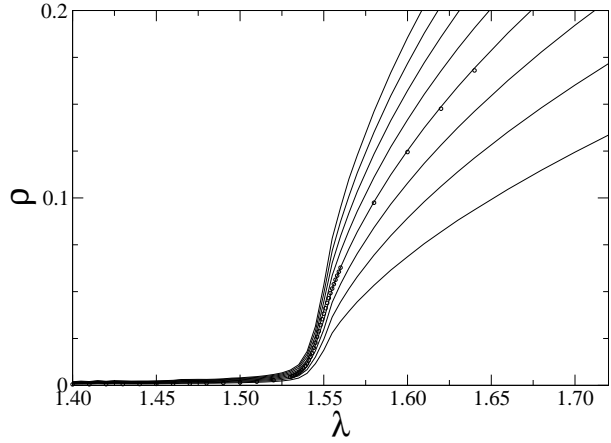}
\end{center}
\caption{QS density of active sites $\rho(q)$ versus $\lambda$, for
sites with $q=3,4,...,10$, from bottom to top.Circles: average over
all sites. System size $L=160$.}
\end{figure}

In Fig. 3 we show the quasistationary density $\rho$ as a function
of the control parameter $\lambda$ for several values of $L$. We
observe, as expected, a continuous phase transition from an active
 to an absorbing state. Since, due to topological constraints,
the Voronoi-Delaunay lattice posses $\overline q \simeq 6$, the
value of the critical point is shifted from $\lambda_c=1.64877(3)$
\cite{dickman99} (regular square lattice) to $\lambda_c=1.54266(4)$
(the increase in $q$ facilitates creation).  This is very close to
the critical value, $\lambda_c = 1.54780(5)$, for the regular
triangle lattice, obtained using the same methods as described
below.  It is notable that the critical value of the disordered
system is about 0.3\% {\it smaller} than that of the regular lattice
with the same average connectivity. Fig. 4 shows how the QS density
of active sites varies with the coordination number $q$.

At the critical point we find that the quasistationary density
decays as a power law, $ \rho \sim L^{-\beta/\nu_\perp}$, as shown
in Fig.5. Our simulation data follow a power law with the exponent
$\beta/\nu_\perp = 0.791(7)$, while the value for DP in two spatial
dimensions is 0.797(3) \cite{dickman99}.

\begin{figure}[!htb]
\begin{center}
\includegraphics[width=8cm,clip=true]{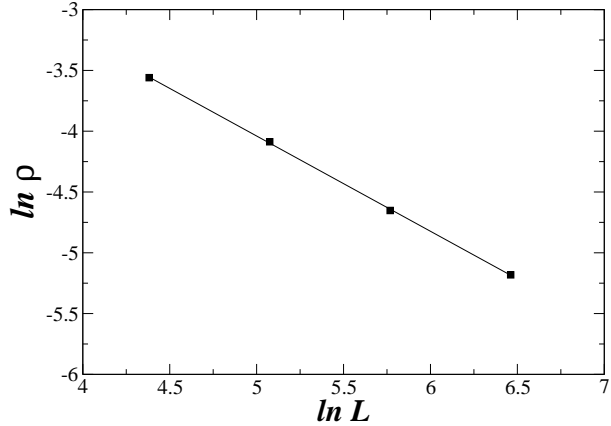}
\end{center}
\caption{QS order parameter $\rho$ versus system size $L$ at
criticality ($\lambda=1.54266$).}
\end{figure}

\begin{figure}[!htb]
\begin{center}
\includegraphics[width=8cm,clip=true]{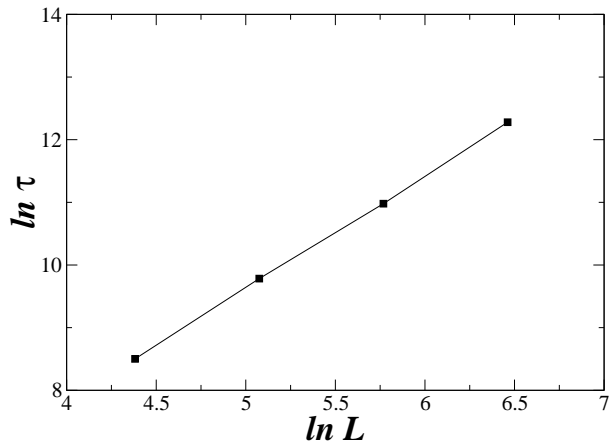}
\end{center}
\caption{Critical lifetime of the QS state $\tau$ versus $L$.}
\end{figure}

Another important quantity is the lifetime of the QS state, $\tau$.
In QS simulations we take $\tau$ to be the mean time between
successive attempts to visit to the absorbing state. Fig.6 shows
that at the critical point, the lifetime also follows a power-law,
$\tau \propto L^{z}$, with $z=\nu_\parallel/\nu_\perp =1.78(3)$, as
compared with the literature value of 1.7674(6) for the DP class
\cite{dickman99}.

Complete characterization of a nonequilibrium universality class
requires the determination of at least three independent critical
exponents. To this end we perform initial decay studies on large
systems, starting with a fully occupied lattice. While the CP with
random dilution exhibits a logarithmic relaxation \cite{adr-dic}, on
the VD lattice we observe a clear power-law decay. Finite size
scaling in this case predicts that $\rho \sim t^{-\delta}$. A
least-squares fit for the data shown in Fig. 7 yields
$\delta=0.453(9)$, in very good agreement with the standard value of
$\delta=0.4523(10)$ for DP \cite{dickman99}.

\begin{figure}[!htb]
\begin{center}
\includegraphics[width=8cm,clip=true]{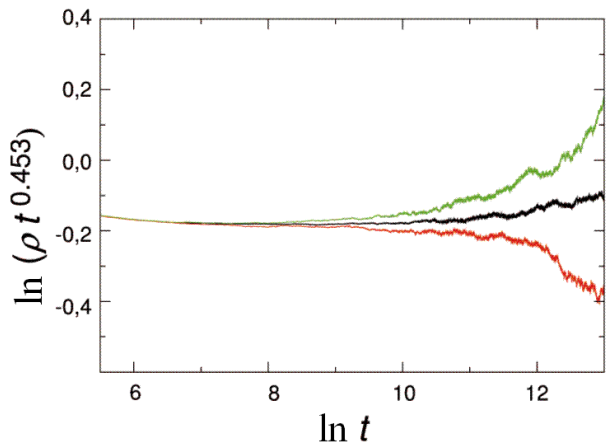}
\end{center}
\caption{Decay of the order parameter starting from a full lattice
of size $L=2560$. $\lambda=1.5428$, $1.5427$ and $1.5426$, from top
to bottom. (Color
  online).}
\end{figure}

Moment ratios (or reduced cumulants) represent an alternative method
for identifying the universality class of a continuous phase
transition \cite{dic-jaf,leite01,oliv-dic06}. Here we analyze the
critical moment ratio $m = \langle \rho^2 \rangle/\langle \rho
\rangle ^2$. This quantity is analogous to Binder's reduced fourth
cumulant \cite{binder}, at an equilibrium critical point: the curves
$m(\lambda,L)$ for various $L$ cross near $\lambda_c$ (the crossings
approach $\lambda_c$), so that $m$ assumes a universal value $m_c$
at the critical point, as can be seen in Fig.8. In this case, our
data yield a universal value of $m_c=1.328(6)$, again in very good
agreement with the best known value for the CP on a regular square
lattice, $m_c=1.3257(5)$ \cite{dic-jaf}.

\begin{figure}[!htb]
\begin{center}
\includegraphics[width=8cm,clip=true]{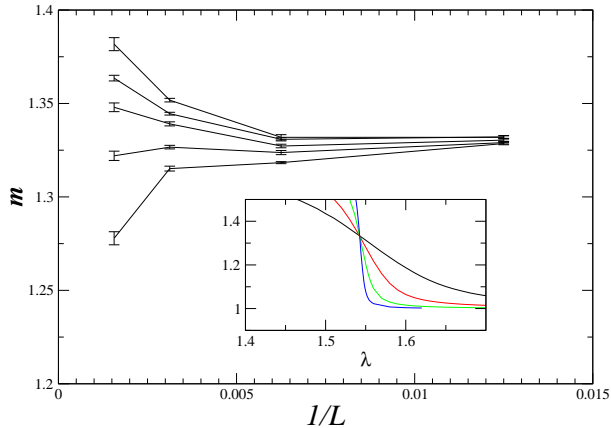}
\end{center}
\caption{Quasistationary moment ratio $m$ versus $\ln L$, for
$\lambda = 1.54256$, $\lambda = 1.54260$,$\lambda =
1.54264$,$\lambda = 1.54268$ and $\lambda = 1.54280$, from top to
bottom. System size: $L=640$. Inset: Quasistationary moment ratio
$m$ versus $\rho$, system sizes: $L=20,40,80,160$. (Color online).}
\end{figure}

In summary, our results reveal that the absorbing phase transition
of the contact process defined on Voronoi-Delaunay random lattice
belongs to the directed percolation universality class. These
results are somewhat surprising, since the wandering exponent for
these lattices was numerically evaluated in a extensive work by
Janke and Weigel \cite{janke}, who found that $\omega=1/2$, i.e, the
relevance criterion for such lattices reduces to the usual Harris
criterion, eq.(1).

In the equilibrium context the Harris-Luck criterion has been
verified numerically on random latices in several models, such as
the Ising  model \cite{janke94, lima00} and percolation
\cite{hsu01}. However, Monte Carlo simulations for the $q=3$ Potts
model \cite{lima00b} as well for the Ising model in 3D
\cite{janke02,lima08} and for the spin-3/2 Blume-Capel model
\cite{lima06} yield results that contradict the relevance threshold
given by the Harris-Luck criterion. Simulation results for
nonequilibrium models, viz. the majority-vote model on a random
lattice \cite{lima05}, also appear to contradict this relevance
criterion.

\begin{figure}[!htb]
\begin{center}
\includegraphics[width=7cm,clip=true,angle=-90]{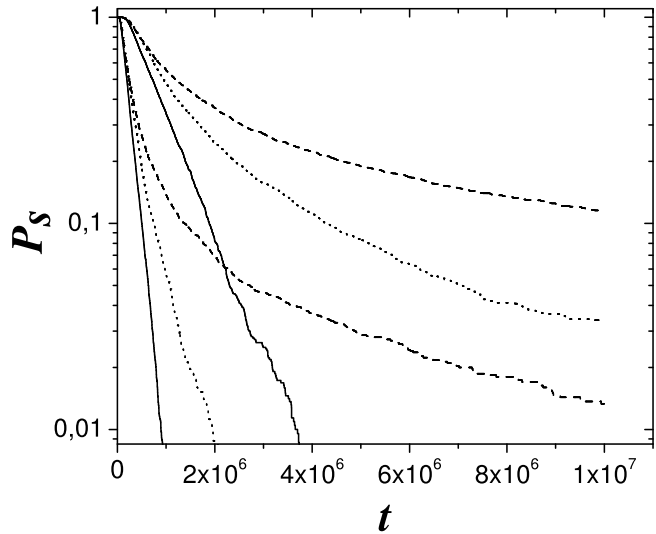}
\end{center}
\caption{Survival probability versus time, in the critical CP on a
Voronoi lattice (solid line), and for the critical CP on a square
lattice with random dilution of 2\% (doted) and 5\% (dashed). System
sizes: $L=640$ (left curves) and $L=1280$ (right).}
\end{figure}

In Refs. \cite{janke,lima08} it is suggested that Voronoi disorder
appears not to alter the critical behavior because it is
intrinsically weak, and that the usual hallmarks of quenched
disorder would in fact manifest themselves in larger systems. In
order to test this hypothesis, we compare in Fig. 9 the survival
probability $Ps$ (starting with a fully occupied lattice) of the CP
on the VD lattice and on a regular (square) lattice with weak
dilution. It is known that the diluted CP exhibits activated
disorder \cite{vojta05}, due to emergence of favorable regions,
leading to logarithmically slow dynamics \cite{adr-dic}. We find
that while in the CP on the VD lattice the survival probability
decays exponentially (as in the ordinary contact process), in the
diluted CP the behavior is clearly different.

Notice that the effect of the ``rare regions" is clearly visible for
the system sizes used here, even for the smallest dilution ($2\%$):
the decay of the survival probability is clearly slower than
exponential. On the square lattice with dilution $x \ll 1$, the
variance of the connectivity $var(q) \simeq 4x$, so that $var(q)
\simeq 0.08$ for dilution $0.02$. This is less than $5\%$ of that
for the VD lattice, where $var(q) = 1.779(2) $. We also should
mention that the effects of quenched disorder in CP are stronger
than in the three-dimensional Ising model: in the latter, quenched
disorder provokes a difference in the second digit in the exponent
$\nu$ \cite{balle98}, while for models in the DP class even weak
disorder changes the critical dynamically properties drastically
\cite{webman}.

\section{Conclusions}

We performed large-scale simulations of the contact process on a
Voronoi-Delaunay random lattice, which exhibits quenched
connectivity disorder in the model. Our results suggest that this
kind of disorder does not alter the DP character of the transition,
in contradiction with the Harris-Luck criterion. Given the large
systems sizes and long simulation times used, it appears unlikely
that the system will eventually cross over to non-DP scaling. Thus
it remains an open question why an argument of the Harris-Luck type
is not applicable in some cases. Our results also reveal that the DP
universality class may be even more robust than asserted in the
usual DP conjecture, in the sense that not all kinds of quenched
disorder are relevant perturbations.

\vspace{1em}

\noindent{\bf Acknowledgments}

This work was supported by CNPq and FAPEMIG, Brazil.


\bibliographystyle{apsrev}

\end{document}